# How did the Suicide Act and Speak Differently Online? Behavioral and Linguistic Features of China's Suicide Microblog Users

[1]Li Guan, [2]Bibo Hao, and [1]Tingshao Zhu

[1]Institute of Psychology, Chinese Academy of Sciences

[2]School of Computer and Control, University of Chinese Academy of Sciences

Author Note

Li Guan is a master candidate in University of Chinese Academy of Sciences majoring applied psychology. His current research interests include the detection and modeling of suicide ideation and well-being via text and behavior in social media.

Bibo Hao is a master candidate in University of Chinese Academy of Sciences majoring computer science. His current research interests include data mining on social media and cyber psychology.

Tingshao Zhu is the director of Division of Social and Engineering Psychology in Institute of Psychology, Chinese Academy of Sciences. His research interest is focused on analyzing user behavior and mental traits via web use.

Corresponding author: Tingshao Zhu

Affiliation: Division of Social and Engineering Psychology, Institute of Psychology, Chinese Academy of Sciences.

Address: Building He-Xie, 16# Lin-Cui Road, Chaoyang District, Beijing, China, 100101

Telephone: (86) 010 6485 1661; Mobile Phone: (86) 150 1096 5509; Email: tszhu@psych.ac.cn

**Abstract**

*Background*: Suicide issue is of great concern in China. Social media provides an active approach to understanding suicide individuals in terms of their behavior and language use. *Aims*: This study investigates how suicide Microblog users in China act and speak differently on social media from others. *Methods*: Hypothesis testing in behavioral and linguistic features was performed between a target group of 33 Chinese Microblog users who have committed suicide and a control group of 30 active users without suicidal ideation. *Results*: Suicide group significantly outnumbered control group in the extent of openly published posts and self-reference, and the intensity of using 7 word categories: negative words/social process words/cognitive process words/emotion process words/negative emotion words/exclusive words/physiological process words. *Limitations*: Information collection and confirmation of suicide users remain difficult. *Conclusions*: It is revealed that suicide people vary from others in certain behavioral and linguistic features in social media. This study fills the niche of suicide studies by noting specified indicators of suicide ideation for Chinese individuals online, providing insights of constructing an online alarm system for early detection and intervention of suicidal individuals.

**Introduction**

**Suicide Issue and Intervention in China**

Suicide issue is as severe in China as in other parts of the world. Nationwide investigation has revealed about 287 thousand suicides and 2 million attempted suicides every year among Chinese citizens; suicide has become the primary cause of life lost for Chinese between the age of 15 to 34 (Beijing Suicide Research and Prevention Center, 2007). The aggregated amount of financial burden and mental trauma resulted from all kinds of fatal and nonfatal self-harm is beyond calculation.

To alleviate suicide damage, detection and intervention of suicide at its early stage is necessary (Awata et al., 2007). Professional assistance from health center is generally available through telephone or face-to-face counseling in mainland China (Li, 2007), and studies for the mental health state of both suicidal individuals and close family members of the suicide have been widely carried out (X. Li & Phillips, 2010; Phillips, Yang, Li, & Li, 2004; Phillips et al., 2009). Nevertheless, current intervention process remains passive and inefficient, as evidence shows that most Chinese individuals with suicidal ideation do not seek professional assistance (Phillips et al., 2002). Moreover, most suicide studies are dependent on surveys, psychological test scales and interviews, faced with restrictions including participants (both participants with suicide ideation and friends/relatives of suicide individuals) conceding their genuine thoughts, difficulty of confirming personal information of the deceased, and ethics issue (Jashinsky et al., 2013).

**Suicide Intervention via Social Media**

Social media provide contemporary experts with a new approach to suicide intervention as more people are expressing their emotions and thoughts in social network platforms. According to Sina Weibo Data Center (2012), more than 100 million posts are posted at Sina Weibo on a daily basis. Contrary to traditional methodology, social media enable scholars to actively search for individuals with suicide ideation, and expose them to assistance in time. For example, minority groups (e.g. teenager lesbians and gay men) in Twitter can be found and offered with help (Silenzio et al., 2009). Individuals reporting suicide plans in Sina Weibo and Facebook also obtained massive amount of comfort and support (Fu, Cheng, Wong, & Yip, 2013; Ruder, Hatch, Ampanozi, Thali, & Fischer, 2011).

Text analysis is a powerful tool for researchers in comprehending thoughts and emotions of Microblog users. People reveal their identity and social status via language (Tausczik & Pennebaker, 2010), whose written form can be transformed to text. As individuals are more willing to leave feelings and thinking on social media due to online disinhibition result (Suler, 2004), and that online texts are consistent with offline texts for suicide individuals (Barak & Miron, 2005), it is practicable to detect the real intentions and potential behaviors of people with suicidal thoughts.

Features of suicidal individuals need to be clarified to advance an identification process. Some previous online suicide case studies have been carried out. For instance, a 13-year-old Chinese boy's blog during the year prior to his suicide was analyzed using the Chinese Linguistic Inquiry and Word Count (CLIWC) program, and posting frequency and language use demonstrated a suicidal process (T. M. Li, Chau, Yip, & Wong, 2014). To increase the knowledge of valid features of suicidal individuals distinguishable from people without suicide ideation, we conduct a benchmark study targeting an "extremely suicidal" group – suicide Microblog users – to discover how they acted and spoke differently from non-suicidal users online.

**Research Objective**

Our research aims to contribute to the suicide studies in the following two aspects. First, as few applicable online criteria of suicide ideation have been noted in previous literature, we intend to discover the strong indicators of suicide ideation by understanding both linguistic and behavioral features of Microblog users who have completed fatal self-harm. Second, as Chinese act and speak differently from people of other countries (Tsai, Simeonova, & Watanabe, 2004), and that Chinese have not been well investigated (Li, Cai, Graesser, & Duan, 2012), we hope to expand the field of suicide studies by portraying the characteristics of China's Microblog users. In the current research, we compare Microblog users who have committed suicide with non-suicidal users in China, combining behavioral and linguistic studies, in the attempt to find distinguishable features for suicide individuals on social media, and to further provide research support to the construction of a social alarm system for suicide ideation.

## Method

**Participants**

**Suicide Users.**

Information of suicide group (19 women, 14 men, $M_{age}$=23.2, age range: 13-48, nationality: China) was collected via cooperation with an officially authenticated Sina Weibo user, who is specialized in the confirmation of identities of deceased Sina Weibo users. Thirty-seven confirmed IDs were provided, four of which were finally eliminated because of the small size of total posts (fewer than twenty posts individually).

**Non-suicidal Users.**

To build a control group of non-suicidal users, 500 Sina Weibo users had been invited to complete an online survey including two scales for detecting suicide ideation: Suicide Possibility Scale (SPS) and Scale for Suicide Ideation (SSI), both in simplified Chinese version. Eligible participants for the control group should meet the following requirements: (1) Pass the screening process of suicide intention detection (see Table 1); (2) Real name authenticated Sina Weibo users with accessible information of age and gender; (3) Have openly published at least 100 posts during the past 1 year. Thirty participants were eventually selected (15 women, 15 men, $M_{age}$=26.7, age range: 21-46, nationality: China). All the available digital records of individuals in both groups were downloaded by calling Application Programming Interfaces (APIs) provided by Sina Weibo. Ethics has been reviewed and granted by the Review Board of Institute of Psychology, Chinese Academy of Sciences.

**Gender and Age Control.**

To eliminate potential influences of feature difference caused by gender and age, T-test was performed to examine the homogeneity of the two groups. There was no significant difference in gender and age between suicide group and control group, $p>0.05$. As a result, revealed feature differences should be attributed to suicide act.

**Behavioral Features of Weibo User**

Nine behavioral features of Sina Weibo users were extracted in terms of their self-expression, interaction with other users, and variation of activeness (See Table 2). The extraction process is based on both research support and data accessibility. For an excluded example, originally there was a tenth feature named as "Activeness of social networking", calculated by the ratio of mutual follow accounts to followed accounts. Despite the richness of previous research suggesting the protective role social network support plays in the ease of suicide ideation (Barak, 2007; Compton, Thompson, & Kaslow, 2005; Kerr, Preuss, & King, 2006), we later discovered that the number of accounts following suicide users generally boomed after media reported the suicide incident. Unable to achieve the data of fan numbers before their suicide act, we had to eliminate this feature because it could not indicate the real social network of suicide users. Ratio data were adopted in most features to eliminate the impact of time discontinuity, since control group remains active online whereas suicide group have no updates any more.

**Linguistic Features of Weibo User**

Linguistic features were specified in Simplified Chinese Microblog Word Count Dictionary (SCMBWC), a Chinese version of Language Inquiry and Word Count (Pennebaker, Francis, & Booth, 2001) which has been demonstrated an effective lexicon for text analysis (Gao, Hao, Li, Gao, & Zhu, 2013). There were altogether 88 features in SCMBWC, covering basic categories in Chinese linguistics such as language process, psychological process, person concern and oral language. TextMind Chinese text analysis system were applied in this study (Gao, Hao, Li, Gao, & Zhu, 2013).

**Statistical Analyses**

Data analysis was performed with SPSS 17.0 using descriptive statistics and independent sample T test to compare the behavioral and linguistic differences between suicide and control group.

## Results

**Descriptive Analysis of Weibo User**

The majority of users in both groups were teenagers or adults below the age of 35, which is consistent with the current age distribution in Sina Weibo as users below the age of 35 took up 90% of the user population (Sina Weibo Data Center, 2013). All participants remained active currently or before their suicide act, as presented in Table 3.

**Group Difference in Behavioral Features**

Table 4 summarizes the comparisons of behavioral features between the two groups. Two features reached statistical significance; suicide group significantly outnumbered control group in the degree of Microblog openness and self-reference. In addition, difference in Microblog transitivity reached marginal significance.

**Group Difference in Linguistic Features**

Significance of difference was revealed in 7 of the 88 linguistic features: negative words/social process words/cognitive process words/emotion process words/negative emotion words/exclusive words/physiological process words, as shown in Table 5. For all the listed features, suicide group outnumbered control group in the word count.

## Discussion

**Strong Indicators of Suicide Ideation in Sina Weibo**

It is revealed in the study that suicide group referred to themselves more frequently than control group. This finding is supported by massive previous research (T. M. Li et al., 2014; Stirman & Pennebaker, 2001; Wang et al., 2013), suggesting that extensive use of first person singular may be one of the most effective indicators of suicide ideation. Individuals inclined to commit suicide are more focused on themselves (Wolf, Sedway, Bulik, & Kordy, 2007), which is also shown in one of the linguistic features, as suicide users expressed more physiological process words (e.g. dizzy, sweat) than non-suicidal users.

According to the study, suicide individuals openly published more Microblogs than common users. In Sina Weibo, a user can set up three levels of accessibility for each post: Level 1: accessible to only to oneself; Level 2: accessible only to some people; and Level 3: accessible to all users. The fact that suicide users generally published more Level 3 posts may suggest a potential intention of exposing their fatal self-harm thoughts. This is partially supported by the test result that they appeared more cognitively and socially active in language use. It has been reported that over 60% suicide individuals had clearly claimed their intention before death (Giovacchini, 1981). As for the current social network environment, most people on social media will offer all kinds of support and help when they find someone reveals strong intention of suicide (Fu et al., 2013; Ruder et al., 2011; Silenzio et al., 2009). There will be great chance that suicidal individuals could be saved if they were identified and exposed to assistance in time.

Previous studies are more concentrated on the richer expression of negative emotions by suicide people (T. M. Li, Chau, Wong, & Yip, 2012; T. M. Li et al., 2014; Pestian et al., 2012), which is echoed by the findings in this study that suicide group generally used more negative emotion words. Nevertheless, it is the first time cognitive words with negative meaning (e.g. no, none, but, except) have been found to represent the suicide ideation. In the Four Factor Model of suicide ideation, it is noted that suicidal individuals show more hostility (Eltz et al., 2007). It can be inferred from the study that hostility may be featured as both emotionally and cognitively negative language use on social media.

**Limitations**

The major obstacle of this study lies in the information collection work of suicide Microblog users. Current notification of suicide users is heavily dependent on media reports and expert confirmation, resulting in relatively small sample size. Due to the anonymity of Internet, collection and confirmation of personal information add up to the difficulty of research. There is possibility that other demographic variables apart from gender and age may become potential disturbance variables in the test (e.g. family income). However, to the best of our effort, age and gender are everything we know of suicide users. To deal with this problem, it is our future work to cooperate with corporations and government to obtain more detailed information.

**Implications**

This study provides insights in the construction of an online detection mechanism for suicide ideation. Although the explanatory power of outstood features from our study in identifying a potentially suicidal Microblog user remains uncertain, we have completed a benchmark study in search of valid indexes to locate individuals in need of suicide intervention promptly. One of the uniqueness of the study is that extreme suicidal ideation are demonstrated by the actual suicide act of users. As a result, any revealed behavioral or linguistic difference between the two groups becomes a more convincing indicator of suicide ideation.

The current research extends online suicide studies to the behavior and context of Chinese individuals. An advantage of the study is that text analysis tool (SCMBWC) developed for simplified Chinese has been adopted to analyze Microblog posts written in Chinese characters. Potential misinterpretations of psychological meanings of the Chinese users caused by translation are avoided (T. M. Li, Chau, Yip, & Wong, 2014). Our study provides support that this tool can be extensively applied in Chinese suicide research field, which is still in its infancy with great research value concerning online methodology.

Table 1

*Scales in the Screener of Suicide Intention Detection*

| Name of Scale | Rules of Scoring | Screening Criteria for No Suicide Ideation |
|---|---|---|
| Scale for Suicide Ideation (SSI) | A subject is considered of no suicide ideation if answer to either Question 4 or 5 is NO | Answer to Q4/Q5 is NO (Beck, Kovacs, & Weissman, 1979) |
| Suicide Possibility Scale (SPS) | Four point Likert Scoring, higher score suggests greater suicide possibility | Score below 50 (Eltz et al., 2007) |

Table 2

*Description of Behavioral Features*

| Name of Feature | Description of Feature |
|---|---|
| Self-description | Character count of user self-description |
| Microblog Openness | The ratio of openly published post count to all post count |
| Microblog Originality | The ratio of original post count to all post count |
| Microblog Transitivity | The ratio of link-embedded post count to all post count |
| Microblog Interaction | The ratio of others-mentioning post count to all post count |
| Group Reference | The average number of first person plural words per post |
| Self-Reference | The average number of first person singular words per post |
| Nocturnal Activeness | The ratio of count of posts sent between 22:00-6:00 to all post count |
| Adoption of Negative Emoticons | The average number of negative emoticons per post |

*Note*. To determine negative emoticons, five psychology master candidates were recruited to evaluate all 1983 Sina Weibo emoticons, and 118 were ultimately confirmed.

Table 3

*Counts of Basic Microblog Records of Participants*

|  | Suicide Group | | Control Group | |
|---|---|---|---|---|
| Microblog Record | M | SD | M | SD |
| Total post | 894.2 | 1105.6 | 1128.8 | 1152.4 |
| Openly published post | 485.0 | 553.8 | 601.8 | 567.0 |
| Favorite page | 198.0 | 434.4 | 174.1 | 390.3 |
| Mutual follow | 81.1 | 64.4 | 165.1 | 110.4 |
| Follow others | 189.5 | 136.2 | 391.6 | 287.1 |

Table 4

*Independent Sample T Test of Behavioral Features*

| Feature Name | Unit | M | | SD | | t | df | p |
|---|---|---|---|---|---|---|---|---|
|  |  | Suicide | Control | Suicide | Control |  |  |  |
| Length of Self-description | Character | 15.09 | 14.60 | 11.63 | 9.88 | 1.09 | 61 | .280 |
| Microblog Openness | Ratio | 0.66 | 0.17 | 0.57 | 0.09 | 2.71 | 61 | .009** |
| Microblog Originality | Ratio | 0.52 | 0.28 | 0.44 | 0.19 | 1.19 | 61 | .241 |
| Microblog Transitivity | Ratio | 0.06 | 0.10 | 0.10 | 0.08 | -1.85 | 61 | .070 |
| Microblog Interaction | Ratio | 0.32 | 0.21 | 0.38 | 0.13 | -1.51 | 61 | .137 |
| Group Reference | Average count per post | 0.08 | 0.06 | 0.06 | 0.02 | .95 | 61 | .345 |
| Self Reference | Average count per post | 0.63 | 0.36 | 0.35 | 0.18 | 3.82 | 61 | .000** |
| Nocturnal Activeness | Ratio | 0.27 | 0.12 | 0.29 | 0.09 | -.69 | 61 | .495 |
| Adoption of Negative Emoticons | Average count per post | 0.10 | 0.13 | 0.08 | 0.10 | .53 | 61 | .495 |

*Note*. *p<0.5, **p<0.01. Units of behavioral features are not consistent, thus were specified in the table.

Table 5

*Word Count as a Percentage of Linguistic Features with Significance of Difference*

| Feature Name (examples of words) | M | | SD | | t | df | p |
|---|---|---|---|---|---|---|---|
|  | Suicide | Control | Suicide | Control |  |  |  |
| Negative words (no, none) | 1.90 | 1.14 | 1.42 | 0.63 | 2.68 | 61 | .009** |
| Social process words (family, greet) | 0.74 | 0.41 | 0.80 | 0.44 | 2.02 | 61 | .048* |
| Emotion process words (grateful, disappoint) | 2.66 | 1.84 | 2.10 | 0.77 | 2.08 | 61 | .044* |
| Negative emotion words (worry, suspect) | 1.04 | 0.56 | 0.96 | 0.34 | 2.71 | 61 | .010** |
| Cognitive process words (understand, question) | 1.40 | 0.92 | 1.14 | 0.46 | 2.17 | 61 | .034* |
| Exclusion words (except, but) | 0.06 | 0.00 | 0.17 | 0.02 | 2.18 | 61 | .033* |
| Physiological process words (dizzy, sweat) | 0.42 | 0.17 | 0.65 | 0.21 | 2.13 | 61 | .040* |

*Note*. *p<0.5, **p<0.01.